\documentclass{article}
\usepackage{spconf,amsmath,graphicx}
\usepackage{algorithm,algorithmic}
\usepackage{epstopdf}
\usepackage{amssymb}
\usepackage{multirow}
\usepackage{epstopdf}
\usepackage{graphicx}
\usepackage{xcolor}
\usepackage{booktabs}
\usepackage{bm}
\usepackage{url}

\def\R{\mathbb{R}}
\def\C{{\mathbb{C}}}

\def\M{{\cal{M}}}
\def\B{{\cal B}}

\DeclareMathOperator*{\argmax}{argmax}

\DeclareMathOperator*{\LUT}{LUT}
\DeclareMathOperator*{\Real}{Real}

\DeclareMathOperator*{\TR}{TR}

\newtheorem{Theorem}{Theorem}

\newtheorem{Definition}{Definition}

\title{Compressed Quantitative MRI: Bloch Response Recovery through Iterated Projection}
\name{Mike Davies$^{\star}$\sthanks{This work was initiated during a research visit by MD to EPFL funded by EPSRC grant EP/K032275/1.} \qquad Gilles Puy$^{\dagger}$ \qquad Pierre Vandergheynst$^{\dagger}$ \qquad Yves Wiaux$^{\ddagger}$}
\address{
$^{\star}$Institute for Digital Communications (IDCom), The University of Edinburgh, EH9 3JL, UK. \\
$^{\dagger}$Institute of Electrical Engineering, Ecole Polytechnique F\'ed\'erale de Lausanne, 1015 Lausanne, CH.\\
$^{\ddagger}$Institute of Sensors, Signals, and Systems, Heriot-Watt University, EH14 4AS, UK.}

\begin{document}
\maketitle

\begin{abstract}
Inspired by the recently proposed Magnetic Resonance Fingerprinting technique, we develop a principled compressed sensing framework for quantitative MRI. The three key components are: a random pulse excitation sequence following the MRF technique; a random EPI subsampling strategy and an iterative projection algorithm that imposes consistency with the Bloch equations. We show that, as long as the excitation sequence possesses an appropriate form of persistent excitation, we are able to achieve accurate recovery of the proton density, $T_1$, $T_2$  and off-resonance maps simultaneously from a limited number of samples.
\end{abstract}
\begin{keywords}
compressed sensing, MRI, Bloch equations, manifolds, Johnston-Linderstrauss embeddings.
\end{keywords}

\section{Introduction}

In the recent paper \cite{Ma-MRF2013}, a new type of MRI acquisition scheme called Magnetic
Resonance Fingerprinting (MRF) is presented for the quantification of multiple tissue properties simultaneously through a single acquisition process. The procedure is composed of 4 key ingredients: (1) the material magnetization is excited through a sequence of random RF pulses; (2) after each pulse the response is recorded through measurements taken from a small portion of $k$-space; (3) a sequence of highly aliased magnetization response images are formed using back projection; and (4) parameter maps (proton density, $\rho$, $T_1$, $T_2$ and off-resonance, $\delta f$) are formed using a bank of matched filters comparing the ``noisy'' magnetization responses for each voxel with the predicted magnetization response for given parameter sets.

Inspired by this technique, we investigate this idea from a compressed sensing (CS) perspective. In \cite{Ma-MRF2013}, it was mentioned that MRF was itself inspired by the recent growth of compressed sensing techniques in MRI, however, the exact link to CS was not made explicit and the paper does not consider a full CS formulation. Indeed the role of sparsity, random excitation and sampling and not clarified. 

Here we identify separate roles for the pulse excitation and the subsampling of $k$-space. We
identify the Bloch response manifold as the appropriate low dimensional signal model on which
the CS acquisition is performed and interpret the ``model-based'' dictionary of \cite{Ma-MRF2013} as a natural discretization of this manifold. We then leverage recent results from \cite{TB-2011} and develop a recovery algorithm with good theoretical guarantees. We conclude with some simulations to demonstrate the efficacy of our approach.

\section{IR-SSFP excitation}

\subsection{The Bloch response manifold}

The MRF process is based upon an Inversion Recovery Steady State Free Precession (IR-SSFP) excitation sequence (see, \emph{e.g.}, \cite{Hargreaves2001, Ganter}). Let $i = 1, \ldots, N$, index the voxels of the imaged slice. We will assume that within each voxel a single isochromat dominates.  The MRF excitation generates a magnetization field that can be observed at each excitation pulse. This field at voxel $i$ at a given time $t$ is a function of the excitation parameters at time $t$ (namely, the flip angle $\alpha_t$ and the repetition time $\TR_t$), the magnetization at time $t-1$, the overall magnetic field, the unknown parameters ${\theta }_i = \{T_1,T_2,\delta f \} \in \M$ associated with the local isochromat, and the voxel's proton density $\rho_i \geq 0$. This overall dynamics for an isochromat can be described by a parametrically excited linear system \cite{Hargreaves2001, Ganter}.
   
Now and subsequently, we will denote the magnetization image sequence by a matrix $X \in \C^{N \times L}$, with $X_{i,t}$ denoting the magnetization for voxel $i$ at the $t^{\text{th}}$ read time. Given that the initial magnetization is known, the magnetization response at any voxel can be written as a parametric nonlinear mapping from $\{\rho_i, {\theta}_i\}$ to the sequence $X_{i,:}$ as follows
\begin{equation}
\label{eq: bloch map}
X_{i,:} = \rho_i B({\bf \theta}_i; \alpha, \TR) \in \C^{1\times L},
\end{equation}
where $L$ is the excitation sequence length, and $B:{ \cal{M}} \rightarrow \C^{1\times L}$ is a smooth mapping induced by the Bloch equation dynamics. Note that we are representing the magnetization response sequence for a given voxel $i$ as $X_{i,:} \in \C^{1 \times L}$, using a Matlab style notation for indexing. Similarly, we will denote the response image at a given time $t$ by the column vector $X_{:,t} \in \C^{N}$.

\subsection{Estimating the Bloch parameters from the responses}
\label{sec: Bloch projection}

In order to be able to retrieve  $\{\rho_i, {\theta}_i\}$, it is necessary that the excitation sequence $(\alpha, \TR)$ is ``sufficiently rich'' so that the voxel's dynamics $X_{i,:}$ is identifiable (random sequences $(\alpha, \TR)$ seem to suffice in practice). Mathematically, this means that there is an embedding of $\R_+ \times \M$ into $\C^{1 \times L}$.\footnote{Strictly speaking, we can only consider this to be an embedding for $\rho_i >0$, otherwise $\theta_i$ is not observable.} We will call $\B = B(\M; \alpha, \TR) \subset \C^{1 \times L}$ the Bloch response manifold and denote its positive cone by $\R_+\B$. 

Inferring $\{\rho_i, {\theta}_i\}$ from the sequence $X_{i,:}$ can be achieved by locating $X_{i,:}$ on $\R_+\B$ and evaluating the associated parameters. This can be approximated in practice by projecting onto the cone of a discretized version of the response manifold. 

First, we take a sufficiently dense discrete sampling of the parameter space $\M$, \emph{i.e.}, $\theta_i^{(k)} = \{T_1^{(k)},T_2^{(k)},\delta f^{(k)} \}_{k = 1:P}$, and then construct a ``dictionary'' $D \in \C^{P \times L}$ of the magnetization responses $D_k =  B(\theta_i^{(k)}; \alpha, \TR)$, $k = 1, \ldots , P$.  We also construct a look-up table (LUT) to provide an approximate inverse for $B(\theta_i; \alpha, \TR)$ such that $\theta_i^{(k)} = \LUT_B (k)$. 

The approximate orthogonal projection onto the cone of response manifold $\R_+\B$, denoted by $\tilde{\cal P}_{\R_+\B}$, satisfies $\tilde{\cal P}_{\R_+\B}(X_{i,:}) = \hat{\rho}_i D_{\hat{k}_i}$, where
\begin{align}
\label{eq: bloch MF}
\hat{k}_i & \in \argmax_{1 \leq k \leq P} \frac{\Real (\langle D_k, X_{i,:} \rangle)}{\|D_k\|_2},
\end{align}
and $\hat{\rho}_i = \max \{\Real ( \langle D_{\hat{k}_i}, X_{i,:} \rangle)/\|D_{\hat{k}_i}\|_2^2, \, 0\}.$ The Bloch parameters corresponding to $X_{i,:}$ satisfies $\hat{\theta}_i = \LUT_B(\hat{k}_i)$.

\section{MRF imaging}

For the complete spatial image, we have ${\bf \theta} \in \M^N$ and $\rho \in \R_+^N$. For convenience, let us denote the full mapping of the product space as $X = f(\rho,\theta)$, with $f: \R_+^N \times \M^N \rightarrow (\R_+\B)^N \subset \C^{N\times L}$. 

Unfortunately, it is impractical to observe the full spatial magnetization $X_{:,t}$ at each repetition time $t$ within the necessary time window. It is necessary to resort to some form of undersampling. We can therefore define the observation sequence $Y_{:,t} \in \C^M$ as
\begin{equation}
\label{eq: observation map}
Y_{:,t} = P(t) F X_{:,t},
\end{equation}
where $F \in \C^{N \times N}$ represents the forward discrete Fourier transform, and $P(t) \in \{0, 1\}^{M \times N}$ is a $t$-dependent projection onto a subset of the output coefficients. We will denote the full linear observation mapping from the spatial magnetization sequence to observation sequence as $Y = h(X)$.

\subsection{MRF Matched filter reconstruction}
\label{sec: MRF reconstruction}

In \cite{Ma-MRF2013}, the image sequence is reconstructed using back projection which is given
by\footnote{This is actually only an approximation since in \cite{Ma-MRF2013} the authors use a nonuniform Fourier transform since their spiral read out does not lie on the DFT grid.}
\begin{equation}\label{eq: MRF image BP}
\hat{X}_{:,t} = F^H P(t)^T Y_{:,t}. 
\end{equation}
Due to the high level undersampling, this process generates extreme aliasing and therefore very noisy images. However, Ma \emph{et al.} argue that by projecting each voxel sequence onto the Bloch response dictionary $D$, the noise can be suppressed and relatively clean parameter maps can be generated. 

The procedure works through a form of noise averaging. Although each individual image is very noisy, the noise is greatly reduced when the voxel sequences are projected onto the Bloch response manifold. However, this ignores the main tenet of CS: aliasing is not noise but interference and under the right circumstances it can be completely removed. We explore this idea next.

\subsection{Compressed Quantitative MRI}
\label{sec: CS-MRF}

In order to be able to retrieve $\{ \rho, {\bf \theta}\}$ from $Y$, we propose a CS solution that has three key ingredients: a random pulse excitation sequence following the original MRF technique; a random subsampling strategy; and an efficient iterated projection algorithm \cite{TB-2011} that imposes consistency with the Bloch equations.

\subsubsection{Bloch response recovery via iterated projection}
 
In \cite{TB-2011}, the Projection Landweber Algorithm (PLA) was proposed as an extension of the popular Iterated Hard Thresholding Algorithm \cite{TB-IHT-2008, TB-IHT-2009}. PLA is applicable to arbitrary union of subspace models as long as we have access to a computationally tractable projection operator within the complete signal space. In our case, the ideal algorithm is given by the recursion
\begin{equation}
X^{(n+1)} = {\cal P}_{(\R_+ \B)^N}\biggl[X^{(n)}+\mu\, h^H \Bigl(Y-h\Bigl(X^{(n)}\Bigr)\Bigr) \biggr],
\end{equation}
where $n$ is the recursion index, ${\cal P}_{(\R_+ \B)^N}$ is the orthogonal projection onto the signal model $(\R_+ \B)^N$, and $\mu$ is a stepsize. 

In practice, we replace the projection ${\cal P}_{(\R_+ \B)^N}$ by the approximate orthogonal projection, denoted by $\tilde {\cal P}_{(\R_+ \B)^N}$, computed by separately projecting the individual voxel sequences $X_{i,:}^{(n)}$ using the projector $\tilde {\cal P}_{\R_+ \B}$ defined in Section \ref{sec: Bloch projection}. We call the resulting algorithm BLIP (BLoch response recovery via Iterated Projection).

The current theory for PLA guarantees stable recovery as long as $h$ satisfies a so-called Restricted Isometry Property (RIP) for the signal model. That is, if there exists a constant $\delta >0$ such that
\begin{equation}
(1-\delta) \| X-\tilde{X}\|_2^2 \leq \tfrac{N}{M}\|h(X-\tilde{X})\|_2^2 \leq (1+\delta) \| X-\tilde{X}\|_2^2 
\end{equation}
for all pairs $X$ and $\tilde{X}$ in $(\R_+ \B)^N$. We will describe how to achieve such an embedding later in section~\ref{sec: excitation and sampling}. Note that the vectors of interest in the above RIP are the \emph{chords} of $\R_+\B$, \emph{i.e.}, the vectors belonging to $U := \{\R_+\B-\R_+\B\}\backslash\{0\}$.

The theory \cite{TB-2011} also requires that $M(1+\delta)/N < 1/\mu < 3M(1-\delta)/(2N)$ for the successful recovery. If $h$ is essentially ``optimal'', \emph{e.g.}, a random ortho-projector, then we should set the stepsize $\mu \approx N/M$ since in the large system limit $\delta \rightarrow 0$. In practice, selection of the correct stepsize is crucial in order to attain good performance from these iterative projection based algorithms \cite{TB-NIHT-2010,TB-2011}. While the matched filter used in \cite{Ma-MRF2013} can be interpreted as a single iteration of PLA with $\mu = 1$, a substantially more aggressive step size is proposed by the theory and results in significant improvements. In practice, we noticed that it is also beneficial to select adaptively the step size for PLA to ensure stability. In the experiments, we use an adaptive stepsize selection as in \cite{TB-NIHT-2010}.

\subsubsection{Strategies for subsampling $k$-space}
\label{sec: excitation and sampling}

In order to enable parameter map recovery, recall that ($1$) the excitation response mapping $f$ must be an embedding (achieved using a random excitation sequence $(\alpha, \TR)$), and ($2$) the sampling operator $h$ must satisfy a suitable RIP. In this section, we show that the last condition is satisfied using an appropriate undersampling strategy, if the vectors in $U$ satisfies certain conditions.

Since we only take a small number of measurements at each repetition time, we cannot expect to achieve a stable embedding without imposing further constraints on the excitation response. For example, if the embedding $f$ was induced in the first few repetition times and all further responses were non-informative, we would not have taken sufficient measurements from the informative portion of the response. Therefore, we need to consider responses that somehow spread the information across the repetition times. We quantify the spread of the information for the excitation response through a \emph{flatness} parameter.
\begin{Definition}
Let $U$ be a collection of vectors $\{u\}$ in $\C^L$. We denote the \emph{flatness} of the these vectors by
\begin{equation}
\lambda := \max_{u \in U} \| u\|_{\infty}/\|u\|_2.
\end{equation}
Note that from standard norm inequalities $L^{-1/2} \leq \lambda \leq 1$.
\end{Definition}

\begin{figure}[t]
\begin{center}
\includegraphics[width=.49\linewidth]{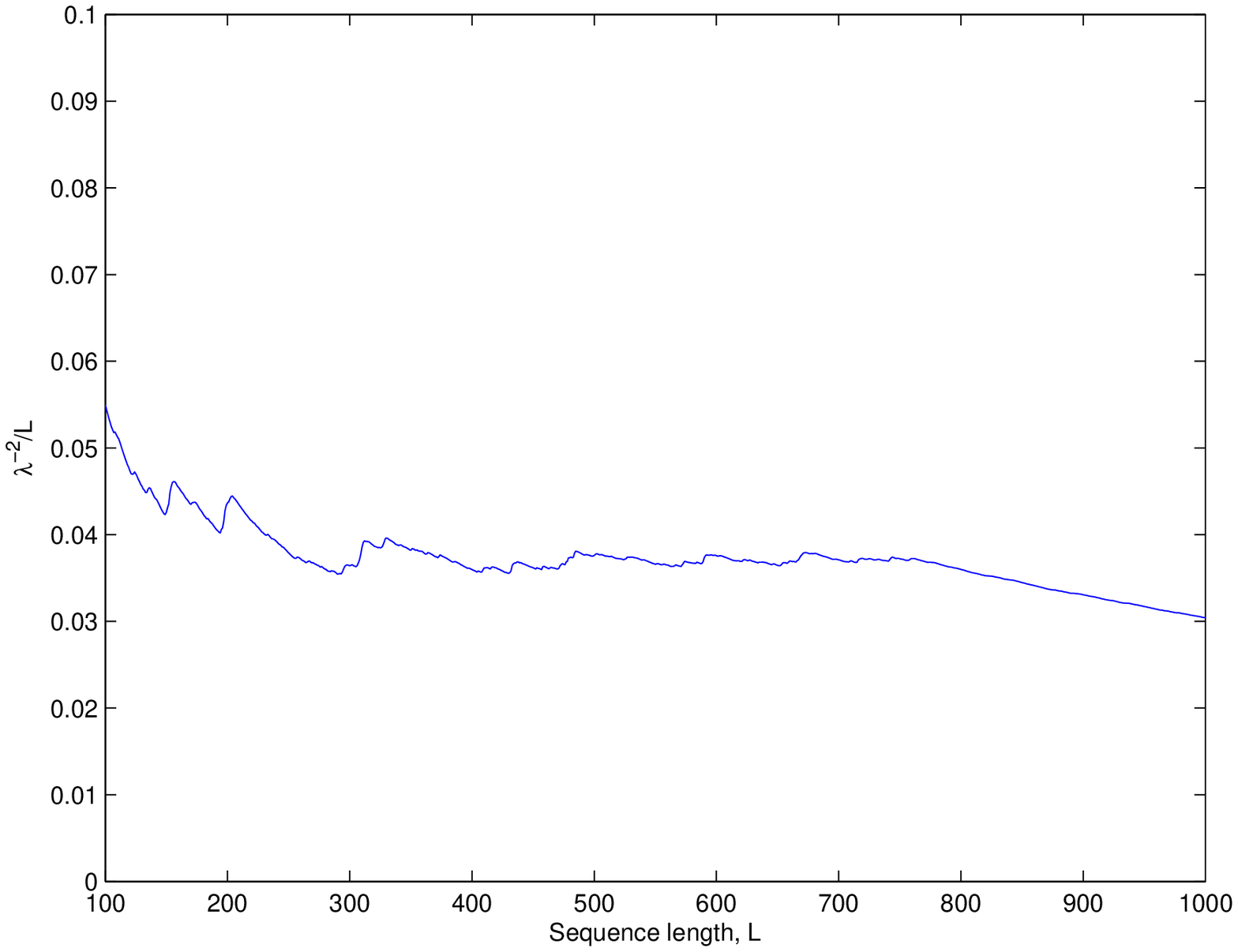}
\includegraphics[width=.49\linewidth]{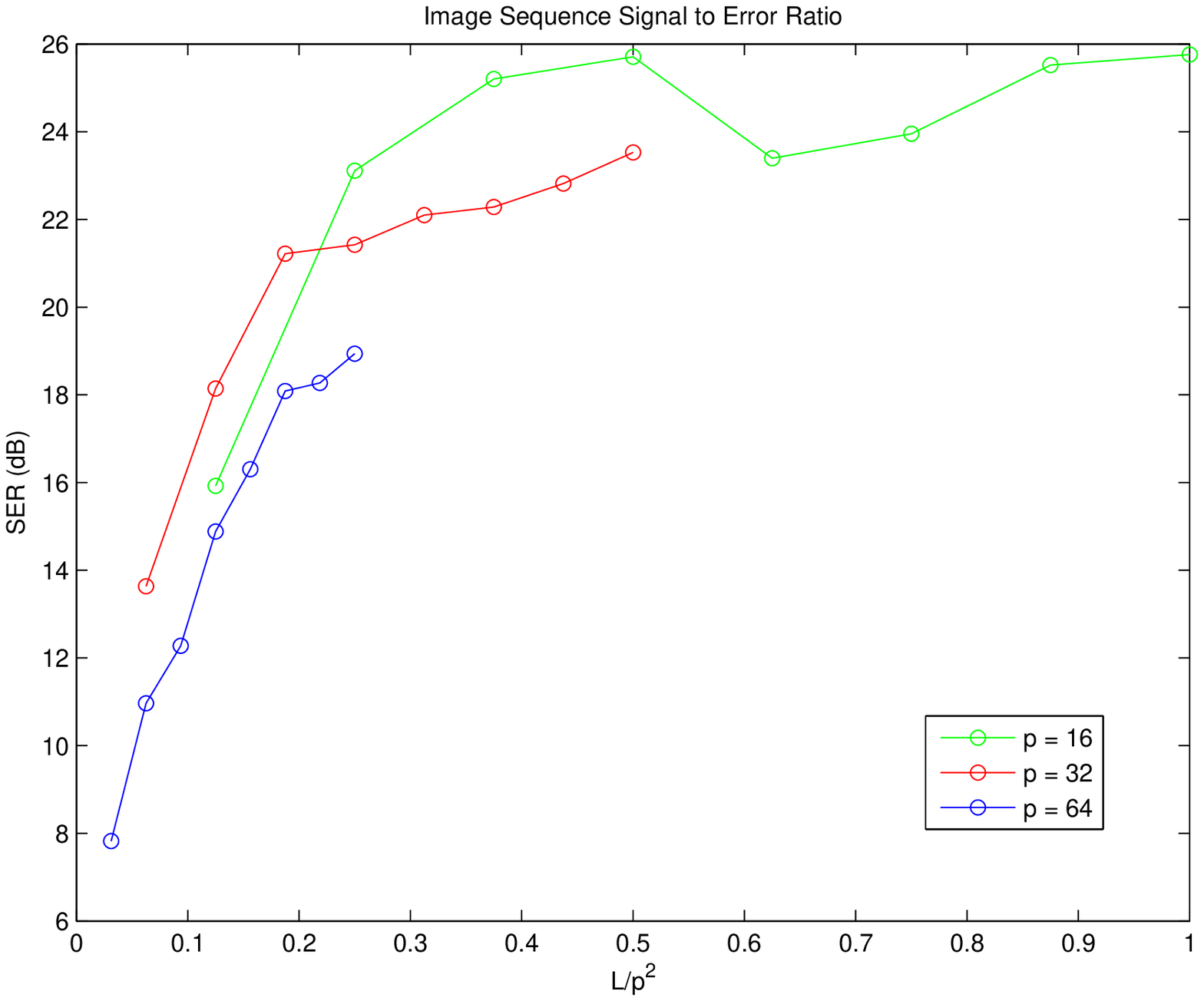}
\caption{Left: $\lambda^{-2}/L$ as a function of $L$ for some example of response differences. Right: plot of the image sequence SER (dB) against $L/p^2$ for three different levels of undersampling $p = 16$ (green), $p = 32$ (red), and $p = 64$ (blue).}
\label{fig: sequence flatness}
\end{center}
\end{figure}

The chords $u \in U$ of $\R_+\B$ are considered sufficiently flat if $\lambda \sim L^{-1/2}$. In Fig.~\ref{fig: sequence flatness}, we present a plot $\lambda^{-2}/L$ as a function of sequence length $L$ for an example of the response differences. From this plot, it can be deduced that $\lambda^{-2}$ grows roughly proportionally to $L$.

In constructing our measurement function $h$, we also note that the signal model contains no spatial structure. Therefore, we should expect to have to uniformly sample $k$-space in order to have the required RIP. Note this is in contrast with the variable density sampling strategy proposed by \cite{Ma-MRF2013}. As no spatial structure is considered, we describe, for simplicity, our sampling strategy for a 1D signal of $N$ pixels. Let $k_1, \ldots, k_N \in \{0, \ldots, N-1\}$ denote the $N$ measurable discrete frequencies. Then, we define a random measurement operator by $P(\zeta_t)F$, where $(\zeta_t)_{1 \leq t \leq L}$ is a sequence of independent random variables uniformly drawn from $\{0, \ldots , p - 1\}$ and $P(\zeta_t) \in \{0,1\}^{M\times N}$ has entries
\begin{equation}
\left(P(\zeta_t)\right)_{i,j} = 
\left\{ 
\begin{array}{ll}
1& \mbox{if~} k_j = (i-1) p +\zeta_t,\\
0& \mbox{otherwise}.
\end{array}               
\right.,
\end{equation} 
with $i = 1, \ldots, M$ and $j = 1, \ldots, N$. For convenience, we assume that $p$ divides $N$ exactly so that $N = pM$. In words, we regularly subsample the $k$-space by a factor of $p$ with random shifts of the selected samples across time. For 2D images, this procedure corresponds to a regular subsampling of the $k$-space in one direction and a complete sampling of the selected lines in the other direction. This can be achieved using a randomized \emph{Echo-Planar Imaging} (EPI) \cite{McKinnon-multi-shotEPI} sequence.

The following theorem shows that random EPI along with an excitation response with appropriate chord flatness is sufficient to provide us with a measurement operator, $h$, that satisfies the RIP on our signal model (proof available in \cite{davies13}).

\begin{Theorem}[RIP for random EPI]
\label{th: randEPI embedding}
Given an excitation response cone $\R_+ \B$ of dimension $d_\B$, whose chords have a flatness $\lambda$, and a random EPI operator $h: (\R_+ \B)^N \rightarrow \C^{M \times L}$. With probability at least $1-\eta$, $h$ is a restricted isometry on $(\R_+ \B)^N - (\R_+ \B)^N$ with constant $\delta$ as long as
\begin{equation}
\lambda^{-2} \geq C \delta^{-2} p^2 d_\B \log ( N /\delta \eta ),
\end{equation}
for some constant $C$ independent of $p, N, d_\B, \delta$ and $\eta$.
\end{Theorem}

Specifically, if $\lambda \sim L^{-1/2}$ then we require $L \sim p^2 d_\B$ excitation pulses. While we might hope to get $L$ of the order of $p d_\B$ it appears that this is not possible, at least for a worst case RIP analysis based on the flatness criterion alone. Indeed, in the experimental section, we will provide evidence to suggest that $L\sim p^2$ is indeed the scaling behavior that we empirically observe.

\section{Experiments}
\label{sec: experiments}

\begin{figure}[t]
\begin{center}
\includegraphics[width=\linewidth]{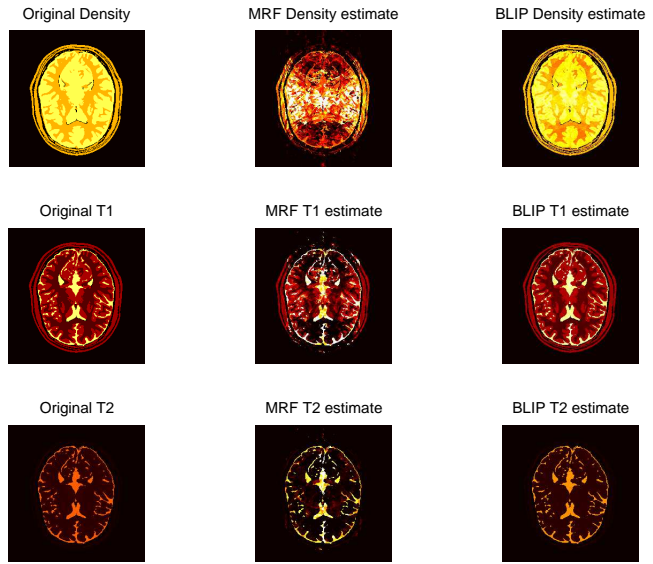}
\caption{\label{fig: visual comparison}
The top row shows the density maps: original (left), MRF estimate (center) and BLIP estimate (right). The middle and bottom rows shows the T1 and T2 maps respectively in the same order.}
\end{center}
\end{figure}
%

\subsection{Setting}

In order to test the efficacy of our method, we performed a set of simulations using an anatomical brain phantom adapted from the anatomical brain phantom of \cite{Collins-1998}, available at the BrainWeb repository.\footnote{\url{http://brainweb.bic.mni.mcgill.ca/brainweb/}} The image contains $256 \times 256$ pixels and was restricted to 6 material components. The material properties were chosen to be representative of the correct tissue type \cite{Hornak-MRIwebbook} and were set so that there is not an exact match to the sampling of the Bloch response manifold. The proton densities were fixed to give little discrimination for individual parameters.

For the excitation sequences, we use IR-SSFP sequences \cite{Ma-MRF2013} with random flip angles drawn from an i.i.d. Gaussian distribution with standard deviation of $10$ degrees. The repetition times were uniformly spaced at an interval of $10$ ms. The Bloch response manifold was sampled in a similar manner to \cite{Ma-MRF2013}, however, we have only considered variation in T1 and T2 here, assuming that $\delta f$ is equal to zero. This results in a dictionary of size $3379\times L$, its range spanning the values for the expected tissue types. For the Fourier subsampling, we use the random EPI sampling scheme detailed in Section~\ref{sec: excitation and sampling}.

\subsection{Results}

To get a visual indication of the performance of the BLIP approach over the original MRF reconstruction, $3$ different parameter estimates for $L = 200$ and $p=16$ are given in Fig.~\ref{fig: visual comparison}. The left hand column shows the ground truth parameter maps while the middle row shows the MRF reconstruction ($1$ iteration of the BLIP algorithm with $\mu = N/M$) and the right hand column shows the BLIP estimates. While the main aspects of the parameter maps are visible in the MRF reconstructions, there are still substantial aliasing artifacts. These are most prominent in the density and $T_1$ estimates. In contrast, the BLIP estimates are virtually distortion-free, indicating that good spatial parameter estimates can be obtained with as little as $200$ excitation pulses.

In the second experiment, we evaluate the image sequence signal-to-error-ratio\footnote{This is calculated as $20 \log_{10}( \|X-\hat{X}\|_F / \|X\|_F)$ for a target signal $X$ with the estimate $\hat{X}$.} (SER) as a function of $L$ and $p$. Recall that the theory suggested that this performance might degrade roughly as a function of $L/p^2$. Fig.~\ref{fig: sequence flatness} shows a plot of the image sequence SER as a function of $L/p^2$ for three different subsampling rates. We can see that the rapid growth of the SER that we associate with successful recovery occurs in each case at roughly the same value of $L/p^2$. This seems to suggest that the predicted scaling behaviour for $L$ and $p$ in random EPI to achieve RIP is of the right order.

\section{Conclusions}

We have presented a principled mathematical framework for compressed quantitative MRI based around the recently proposed technique of MRF \cite{Ma-MRF2013}. The key elements of our approach have been: the characterization of the signal model through the Bloch response manifold; the identification of a provably good reconstruction algorithm based on iterative projection; an excitation response condition based on a newly introduced measure of \emph{flatness}; and a random EPI scheme that has the necessary RIP condition.

While the current work is targeted at a CS framework for MRF, we believe that many elements of it should be more broadly applicable. Specifically, the RIP condition for randomized EPI may have applications in other MR imaging strategies, and the characterization of excitation response in terms of flatness could be a useful tool for the analysis of other CS schemes involving some form of active sensing.

\bibliographystyle{IEEEbib}
\bibliography{biblio.bib}

\end{document}